\documentclass[journal]{IEEEtran}

\usepackage[T1]{fontenc}
\usepackage{graphicx}
\usepackage[caption=false]{subfig}
\usepackage{mathrsfs,amsmath,amsfonts,amssymb,amsthm,amstext,amscd,amsxtra,amsopn}
\usepackage{diagbox}
\usepackage{lettrine}
\usepackage{cite}
\usepackage{eucal}
\usepackage{soul}
\providecommand{\keywords}[1]{\textbf{\textit{Index terms---}} #1}
\usepackage{bm}
\usepackage{multirow}
\usepackage{esint}
\usepackage{empheq}
\usepackage{comment}
\usepackage{boxhandler}
\usepackage{bm}
\usepackage{algorithm}
\usepackage[noend]{algpseudocode}
\usepackage{etoolbox}
\apptocmd{\sloppy}{\hbadness 10000\relax}{}{}
\usepackage{leftidx}
\usepackage{makecell}
\usepackage{stackrel}
\usepackage{xcolor}
\usepackage{hyphenat}
\usepackage{nicefrac}
\usepackage[normalem]{ulem}
\usepackage[hyphens]{url}
\usepackage[hidelinks]{hyperref}
\hypersetup{breaklinks=true}
\renewcommand{\Re}{\mathrm{Re}\,}

\renewcommand{\vec}[1]{\bm{#1}}

\newcommand{\matvec}[1]{\mathbf{#1}}

\algnewcommand{\IfThenElse}[3]{
  \State \algorithmicif\ #1\ \algorithmicthen\ #2\ \algorithmicelse\ #3}
\algnewcommand{\IfThen}[3]{
  \State \algorithmicif\ #1\ \algorithmicthen\ #2}

\DeclarePairedDelimiter\abs{\lvert}{\rvert}
\newcommand\norm[1]{\left\lVert#1\right\rVert}

\newcommand\Aop{{\CMcal{A}}}

\newcommand\Gop{{\CMcal{G}}}

\newcommand\Oop{{\CMcal{O}}}
\newcommand\Pop{{\CMcal{P}}}

\newcommand\Xop{{\CMcal{X}}}
\newcommand\Zop{{\CMcal{Z}}}

\newcommand\At{{\cal{A}}}

\newcommand\Kt{{\cal{K}}}

\newcommand\Nt{{\cal{N}}}

\newcommand{\pluseq}{\mathrel{+}=}

\def\CC{{C\nolinebreak[4]\hspace{-.05em}\raisebox{.15ex}{\small\bf ++}}}

\makeatletter
\def\BState{\State\hskip-\ALG@thistlm}

\newenvironment{breakablealgorithm}
  {
   \begin{center}
     \refstepcounter{algorithm}
     \renewcommand{\caption}[2][\relax]{
       {\raggedright\textbf{\ALG@name~\thealgorithm :} ##2\par}%
       \kern2pt\hrule\kern2pt\hrule\kern2pt
     }
  }{
     \kern2pt\hrule\kern2pt\hrule\kern2pt\relax
   \end{center}
  }

\makeatother

\algnewcommand{\Initialize}[1]{%
  \State \textbf{Initialize:}
  \Statex \hspace*{\algorithmicindent}\parbox[t]{.8\linewidth}{\raggedright #1}
}

\newcommand{\iu}{{\mathrm{i}}}

\makeatletter
\DeclareRobustCommand{\pder}[1]{%
  \@ifnextchar\bgroup{\@pder{#1}}{\@pder{}{#1}}}
\newcommand{\@pder}[2]{\frac{\partial#1}{\partial#2}}
\makeatother

\title{Compression of volume-surface integral equation matrices via Tucker decomposition for magnetic resonance applications}

\author{Ilias I. Giannakopoulos, ~\IEEEmembership{Member,~IEEE}, Georgy D. Guryev, Jos{\'e} E. C. Serrall{\'e}s, ~\IEEEmembership{Member,~IEEE}, Ioannis P. Georgakis, Luca Daniel, ~\IEEEmembership{Member,~IEEE}, Jacob K. White, ~\IEEEmembership{Fellow,~IEEE}, Riccardo Lattanzi, ~\IEEEmembership{Senior Member,~IEEE}

\thanks{This work was supported by NIH R01 EB024536 and by NSF 1453675. It was performed under the rubric of the Center for Advanced Imaging Innovation and Research (CAI$^2$R, www.cai2r.net), a NIBIB Biomedical Technology Resource Center (NIH P41 EB017183).}
\thanks{Ilias I. Giannakopoulos, Ioannis P. Georgakis and Riccardo Lattanzi are with Center for Advanced Imaging Innovation and Research (CAI$^2$R), Department of Radiology, New York University Grossman School of Medicine, NY, USA.}
\thanks{Georgy D. Guryev, Jos{\'e} E.C. Serrall{\'e}s, Luca Daniel, and Jacob K. White are with the Research Laboratory of Electronics, Department of Electrical Engineering and Computer Science, Massachusetts Institute of Technology, Cambridge, MA, USA.}
\thanks{Riccardo Lattanzi is also with the Bernard and Irene Schwartz Center for Biomedical Imaging, Department of Radiology, New York University Grossman School of Medicine, NY, USA and the Vilcek Institute of Graduate Biomedical Sciences, New York University Grossman School of Medicine, NY, USA.}}

\begin{document}
\bstctlcite{IEEEexample:BSTcontrol}

\maketitle

\begin{abstract}
In this work, we propose a method for the compression of the coupling matrix in volume\hyp surface integral equation (VSIE) formulations. VSIE methods are used for electromagnetic analysis in magnetic resonance imaging (MRI) applications, for which the coupling matrix models the interactions between the coil and the body. We showed that these effects can be represented as independent interactions between remote elements in 3D tensor formats, and subsequently decomposed with the Tucker model. Our method can work in tandem with the adaptive cross approximation technique to provide fast solutions of VSIE problems. We demonstrated that our compression approaches can enable the use of VSIE matrices of prohibitive memory requirements, by allowing the effective use of modern graphical processing units (GPUs) to accelerate the arising matrix\hyp vector products. This is critical to enable numerical MRI simulations at clinical voxel resolutions in a feasible computation time. In this paper, we demonstrate that the VSIE matrix\hyp vector products needed to calculate the electromagnetic field produced by an MRI coil inside a numerical body model with $1$ mm$^3$ voxel resolution, could be performed in $\sim 33$ seconds in a GPU, after compressing the associated coupling matrix from $\sim 80$ TB to $\sim 43$ MB. 
\end{abstract}


\section{Introduction} \label{sc:I}
\IEEEPARstart{M}{agnetic} resonance (MR) imaging (MRI) provides high\hyp resolution images of the interior anatomical and physiological structure of the human body, with exquisite soft\hyp tissue contrast. The quality of MR images, as well as the achievable spatial and temporal resolution, depend on the available signal\hyp to\hyp noise ratio (SNR). SNR increases with the main magnetic field strength. This fact motivated the recent development of $7$ Tesla (T) clinical MR scanners and research\hyp only scanners with field strengths as high as $11.7$ T \cite{anderson2011effect}. At ultra\hyp high\hyp field (UHF) MRI ($\ge 7$ T), the radio frequency (RF) wavelength is short. This results in strong interactions between biological tissues and the electromagnetic (EM) field generated by the RF coils \cite{jin1997sar, Lattanzi2009, zhang2013complex, cosottini2014short}. Such interactions could compromise image quality and patient safety. To address these issues, EM modeling is often used to predict and manipulate the EM field distribution during RF coil design.
\par
Integral equation (IE) methods are suitable options for EM analysis in MRI. First, they do not suffer from grid dispersion errors \cite{Taflove1989, Lee1992}, in contrast with the finite\hyp difference\hyp time\hyp domain (FDTD) and finite\hyp element\hyp methods (FEM), because the Green's functions in the underlying IE formulation act as an exact EM field propagator from a source to an observation point. Second, for the case of single\hyp frequency problems, IE algorithms can be extensively customized with the use of numerical linear algebra techniques for fast and accurate simulations, tailored to specific applications \cite{phillips1997precorrected, tambova2017generalization, giannakopoulos2019memory, Villena2016}.
\par   
For example, the MAgnetic\hyp Resonance Integral Equation (MARIE) suite \cite{Villena2016, Polimeridis2014} was developed to numerically compute the EM field distribution generated by RF coils in the human body during MRI. MARIE combines surface and volume integral equations (SIE,VIE), employing a triangular tessellation for the RF coils' conductors and a uniform voxelized grid discretization for the body models. RWG basis functions \cite{Rao1982} and polynomial basis functions \cite{Polimeridis2014, georgakis2019fast} are used to compute the unknowns of the surface and volume IE, respectively. Matrix\hyp vector products are accelerated using the fast Fourier transform (FFT). 
\par
The VIE computational engine of MARIE has been recently employed for the forward problem in Global Maxwell Tomography (GMT) \cite{serralles2019noninvasive}, a technique that iteratively solves an ill\hyp conditioned inverse problem to extract electrical properties from volumetric MR measurements. In the first experimental demonstration of GMT with a uniform phantom, constant incident fields were used for all iterations \cite{serralles2019noninvasive}. More recently, it was shown in simulation that GMT could accurately reconstruct brain electrical properties at $7$ T using a tailored RF coil array \cite{giannakopoulos2020magnetic}. However, in order to confirm this with in\hyp vivo experiments, the currents on the coil conductors cannot be simply considered constant as in the initial experiment with a uniform phantom. Instead, the incident fields must be updated at each iteration of GMT to account for changes in the sample electrical properties distribution. Therefore, GMT must be implemented with a volume\hyp surface IE (VSIE) framework, in which a coupling matrix represents the coil\hyp body interactions in the IE system of equations \cite{Villena2016}. Such approach requires a large amount of memory, which could prevent using clinically\hyp relevant voxel resolutions and fine coil meshes.
\par
The aim of this work is to use Tucker decomposition \cite{Tucker1966} to perform a column\hyp wise compression of the VSIE coupling matrix, in order to limit the associated memory demand and enable the computation of the relevant matrix\hyp vector products in GPUs. Our approach was motivated by previous work \cite{giannakopoulos2019memory} on the reduction of the memory footprint of FFT\hyp based VIE Green's function tensors and the acceleration of matrix\hyp vector products in VIE using GPU. Tucker decomposition belongs to a larger family of tensor decompositions and have been used successfully in the past for matrix compression inside IE-based simulations for EM applications. Examples include EM simulations of realistic body models simulations for UHF MRI \cite{Giannakopoulos2018, giannakopoulos2019memory} and capacitance extraction \cite{zhang2016tucker, wang2020voxcap, qian2020compression}. Other tensor decompositions could be used \cite{Oseledets2011, Khoromskij2010, Grasedyck2010}, but for the intrinsic 3D nature of the problem at hand, Tucker optimizes operations and memory complexity. In cases where the coil is placed far from the scatterer, the coupling matrix can be first compressed with a 2D cross approximation method \cite{Tyrtyshnikov1996a, Tyrtyshnikov1996b, Goreinov1997} and then further compressed by applying our proposed technique to the resulting matrices. Towards this direction, we developed an algorithm based on the adaptive cross approximation (ACA) \cite{Kurz2002, Bebendorf2003} to efficiently perform our compression approach within the iterative loop of ACA. 
\par
Other memory\hyp friendly techniques are available for the fast implementation of matrix\hyp vector products in VSIE simulations: the magnetic resonance Green's function (MRGF), the fast multipole method (FMM), the precorrected FFT method (pFFT), and the Multilevel Nonuniform Grid Algorithm (MNGA). The MRGF \cite{Villena2016} is a model order reduction technique that can considerably accelerate the solutions of the VSIE system. However, the required computational time can be overwhelming when fine voxel resolutions and PWL basis functions are used. The FMM \cite{greengard1987fast, Coifman1993, shanker2007accelerated} has been extensively used for the compression of the Method of Moments (MoM) matrix appearing in IEs and could proven to be a good alternative for solving the VSIE system studied herein. Nevertheless, in the presented work, we are only interested in the compression of an off\hyp diagonal block of the full MoM matrix (i.e., the coupling matrix), since the remaining blocks can be handled efficiently with other methods presented in \cite{Rao1982, Polimeridis2014, giannakopoulos2019memory}. The pFFT method \cite{phillips1997precorrected} could be used to project the discretized coil's elements onto an extended VIE domain, where the Green's function tensors are compressible with the Tucker decomposition (pFFT+Tucker) \cite{giannakopoulos2019memory}. However, this approach would be effective only when the coil is close to the scatterer, like for the close-fitting coil studied in section IV.B.1 \cite{guryev2019fast}. In fact, in such situation the extended VIE domain would be larger than the original one by only a few voxels in each direction, allowing the matrix\hyp vector products to fit in a GPU. As a result, the pFFT+Tucker approach could be more efficient than our proposed method for such geometries, although more complex to implement. Finally, the MNGA \cite{brick2009multilevel} aims to accelerate the solutions of the MoM system of equations through interpolation of sampled fields generated by source distributions in spherical grids. However, this method works well for quasiplanar and elongated geometries, with dimensions larger than the operating wavelength, which is not the case in UHF MRI. While the previously described methods could be applied to VSIE simulations, since we are interested in problems discretized on a uniform 3D grid for UHF MRI frequencies, in this work we chose to explore approaches based on Tucker decomposition that allows us to exploit the low\hyp rank properties encoded in the VSIE coupling matrix.
\par
The remainder of this paper is organized as follows. In Section II, we summarize the relevant technical background and equations related to the standard VSIE method. We also show, as an example application, the use of the VSIE coupling matrix in the GMT inverse problem formulation. In addition, we outline the Tucker decomposition and the ACA method. In Section III, we introduce the Tucker\hyp based assembly technique of the coupling matrix along with the novel algorithms for the matrix\hyp vector products implementation. Moreover, we present a new algorithm for a memory friendly assembly of the coupling matrix, based on the combination of ACA with the Tucker decomposition. Section IV describes a series of numerical experiments aimed at investigating the multilinear rank and the memory requirements of the compressed matrix in different scenarios, along with time footprint of the matrix\hyp vector product for various mesh discretizations. Section V discusses the results. Section VI summarizes the work and provides a few take home points. The following TABLE I lists the notations used in this work.

\begin{table}[ht]
\caption{Notation} \label{tb:n1} \centering
\begin{tabular}{c| l }
\hline\hline\\[-0.4em]
Notation                 	  				 & Description                                         \\[0.4em] \hline \\[-0.4em]
$a$                      	  				 & Scalar                                 	           \\[0.3em]
$\vec{a}$                	  				 & Vector in $\mathbb{C}^3$                 		   \\[0.3em]
$\matvec{a}$              	  				 & Vector in $\mathbb{C}^{n}$                          \\[0.3em]
$A$                      	  				 & Matrix in $\mathbb{C}^{n_1 \times n_2}$             \\[0.3em]
$A^T$                    	  				 & Transpose of matrix                                 \\[0.3em]
$A^*$                    	  				 & Conjugate transpose of matrix                       \\[0.3em]
$\Aop$                        			  	 & Tensor in $\mathbb{C}^{n_1 \times n_2 \times n_3}$  \\[0.3em]
$\mathscr{A}$                      	  	     & Struct containing Tucker compressed tensors         \\[0.3em]
$\At$              	          				 & Operator acting on vectors in $\mathbb{C}^3$        \\[0.3em]
$\times_i$                                   & n-mode products                                     \\[0.3em]
$\iu$                                        & Imaginary unit $\iu^2 = -1$                         \\[0.3em]
\hline\hline
\end{tabular}
\end{table}

\section{Technical Background} \label{sc:II}

\subsection{Volume-Surface Integral Equation} 

\subsubsection{Coupled Linear System}

Let us consider an IE\hyp based solver for the EM wave analysis in MRI applications. The body domain $\Omega$ can be modeled with the current\hyp based VIE (JVIE), while the conducting surfaces (receive and transmit RF coil arrays, coil shields, and gradient coils) with the SIE. The IEs are solved with the MoM \cite{harrington1993field}. The resulting system of equations can be written in the following block matrix form.
\begin{equation}
\begin{bmatrix}
Z_{\rm cc} & T^T_{\rm bc} \\
T_{\rm bc} & Z_{\rm bb}
\end{bmatrix} \begin{bmatrix}
J_{\rm s} \\
J_{\rm p} 
\end{bmatrix} = \begin{bmatrix}
V \\
0
\end{bmatrix}.
\label{eq:n1}
\end{equation} 

Here, $Z_{\rm cc} \in \mathbb{C}^{m \times m}$ is the Galerkin matrix that models interactions between the coil's discretization elements (triangles' common edges), with the aid of the free\hyp space Green's function that appears in the SIE formulation. It associates the equivalent surface currents on the coil ($J_{\rm s} \in \mathbb{C}^{m \times p}$), with the voltage excitation matrix $V \in \mathbb{C}^{m \times p}$. The coil conductors are modeled with triangular meshes, and the unknown surface equivalent currents are approximated with RWG basis functions \cite{Rao1982}. $m$ is the number of RWG, or non\hyp boundary, triangle edges that appear on the mesh, whereas $p$ is the number of excitation ports (i.e., number of coil's channels) of the conducting surfaces. 
\par
The matrix $Z_{\rm bb} \in \mathbb{C}^{qn_{\rm v} \times qn_{\rm v}}$ is another Galerkin matrix which models the EM interactions of an external volumetric EM field, produced by the coil, with the body. Specifically, the matrix relates the polarization currents in $\Omega$ to an external incident EM field produced by the conducting surfaces. $n_{\rm v}$ is the number of voxels and $q$ the number of basis functions per voxel. Differently than $Z_{\rm cc}$, $Z_{\rm bb}$ requires a large amount of memory, even for coarse resolutions. To handle that, $\Omega$ can be discretized over a voxelized uniform grid, giving $Z_{\rm bb}$ a three\hyp level Block\hyp Toeplitz with Toeplitz Blocks (BTTB) structure. As a result, only the defining columns of the BTTB matrix need to be stored and the matrix\hyp vector product can be accelerated using the FFT, as in \cite{Catedra1989, Zwamborn1992, Gan1994, Jin1996, Beurden2008, Markkanen2012, Oijala2014, Polimeridis2014}. The unknown polarization currents ($J_{\rm p} \in \mathbb{C}^{qn_{\rm v} \times p}$) can be discretized with polynomial basis functions, either piecewise constant \cite{Polimeridis2014} (PWC, $3$ unknowns per voxel) or piecewise linear \cite{georgakis2019fast} (PWL, $12$ unknowns per voxel), and a single\hyp voxel support.
\par
The presence of conductive tissue near the coil conductors perturbs $J_{\rm s}$ from their values in free\hyp space. In fact, the voltage excitations at the coil's ports create incident EM fields that scatter from the dielectric body back to the coil conductors, changing their current distribution. The coupling matrix $T_{\rm bc} \in \mathbb{C}^{q n_{\rm v} \times m}$ is used to account for this effect, by modeling the coupling interactions between the dyadic Green's function \cite{tai1994dyadic} of the SIE and the VIE formulations. Specifically, in equation (\ref{eq:n1}), $T_{\rm bc}$ maps electric surface equivalent currents to electric fields through the $\Nt$ Green's function operator of VIE:  
\begin{equation}
\Nt \left(\vec{s}\right) \triangleq \nabla \times \nabla \times \int\limits_{\Omega} g\left(\vec{r}-\vec{r}'\right) \vec{s}\left(\vec{r}'\right)d^3\vec{r}'.
\label{eq:nNt}
\end{equation}
$g$ is the free\hyp space Green's function, or fundamental Helmholtz solution, and it is equal to
\begin{equation}
g\left(\vec{r} - \vec{r}' \right) = \frac{e^{-\iu k_0 \abs{\vec{r}-\vec{r}'}}}{4 \pi \abs{\vec{r}-\vec{r}'}},
\label{eq:n6}
\end{equation}
where $k_0$ is the free-space wavenumber, $\vec{r}$ the source point, and $\vec{r}'$ the observation point. Each element of the VSIE coupling matrix is a 5D integral formed from the inner product between the discretized $\Nt$ operator applied on a VIE basis function, and an RWG basis function.       

\subsubsection{VSIE implementation of GMT} 

GMT estimates tissue electrical properties from MR measurements by solving an inverse problem \cite{serralles2019noninvasive}. In GMT, the cost function compares actual measurements against simulated measurements of the relative $b_1^+$ fields generated by multiple sources (e.g., multiport transmit coils) inside a sample and iteratively updates the estimate of the sample's electrical properties. GMT was initially demonstrated using the JVIE formulation for the solutions of the forward problem, therefore ignoring the effect of the dielectric sample on the incident fields. However, these interactions must be taken into account for accurate in\hyp vivo experiments with close\hyp fitting RF coils. In other words, the GMT framework must be ported from a VIE to a VSIE formulation, in which the incident fields are not constant but calculated at each GMT iteration as
\begin{equation}
\begin{aligned}
E^{\rm inc}\left( \matvec{\epsilon_r} \right) &= T_{\rm bc} J_{\rm s}\left( \matvec{\epsilon_r} \right),\\
H^{\rm inc}\left( \matvec{\epsilon_r} \right) &= Z_{\rm bc} J_{\rm s}\left( \matvec{\epsilon_r} \right).
\end{aligned}
\label{eq:n3}
\end{equation}
where $E^{\rm inc}$ and $H^{\rm inc}$ are the discretized incident electric and magnetic fields respectively. $\epsilon_r$ is the complex permittivity. $Z_{\rm bc}$ maps the equivalent surface electric currents to the magnetic fields with the aid of the $\Kt$ operator:
\begin{equation}
\Kt \left(\vec{s}\right) \triangleq \nabla \times \int\limits_{\Omega} g\left(\vec{r}-\vec{r}'\right) \vec{s}\left(\vec{r}'\right)d^3\vec{r}'.
\label{eq:nKt}
\end{equation}
In addition, in the new implementation, the gradient of the GMT's cost function will require to solve a Hermitian adjoint system of equations that includes multiplications with the conjugate transpose of $Z_{\rm bc}$. 
\par
Matrix\hyp vector products involving the coupling matrix are typically performed without storing the full matrix, due to its intractably large size. In the case of iterative inverse problem solutions, such as in GMT, this approach could considerably increase the computation time, because it requires the re-assembly of the full matrix at each iteration. In the next sections, we propose a compression algorithm that reduces the computational burden, by enabling one to assembly the full coupling matrix only once and then perform just the matrix\hyp vector multiplications in each of GMT's iterations.

\subsection{Numerical Linear Algebra Methods}

\subsubsection{Tucker Decomposition}

A 3D tensor $\Aop \in \mathbb{C}^{n_1 \times n_2 \times n_3}$ can be decomposed with the Tucker model \cite{Tucker1966} in the following form:
\begin{equation}
\begin{aligned}
\Aop &= \Gop \times_1 U^1 \times_2 U^2 \times_3 U^3, \: \text{or} \\
\Aop_{ijk} &= \sum_{\chi=1}^{r_1} \sum_{\psi=1}^{r_2} \sum_{\zeta=1}^{r_3} \Gop_{\chi \psi \zeta} U^{1}_{i \chi} U^{2}_{j \psi} U^{3}_{k \zeta}.
\end{aligned}
\label{eq:Tucker}
\end{equation}
Here $U^{\gamma} \in \mathbb{C}^{n_{\gamma} \times r_{\gamma}}, \gamma = 1,2,3$, are unitary matrices, dubbed as Tucker factors, while $\Gop \in \mathbb{C}^{r_1 \times r_2 \times r_3}$ is the Tucker core. The dimensions of the Tucker core indicate the multilinear (or Tucker) ranks of $\Aop$. The symbols $\times_{\gamma}$ are called $n$\hyp mode products and perform a convolution over the $\times_{\gamma}$ axis, for example, the $\times_1$ product performs the following operation:
\begin{equation}
\Pop = \Gop \times_1 U^1 \Leftrightarrow \Pop_{i \psi \zeta} = \sum\limits_{\chi=1}^{r_1} \Gop_{\chi \psi \zeta}U^1_{i\chi}. 
\end{equation}
Here, $\Pop \in \mathbb{C}^{n_1 \times r_2 \times r_3}$. The expansion of $\Aop$ in equation (\ref{eq:Tucker}) can be truncated to a desired tolerance, and return an approximation of $\Aop$. A visual representation of Tucker decomposition can be seen in Fig. 1.

\renewcommand{\thefigure}{1}
\begin{figure}[ht!]
\begin{center}
\includegraphics[width=0.48\textwidth, trim={0 0 0 0}]{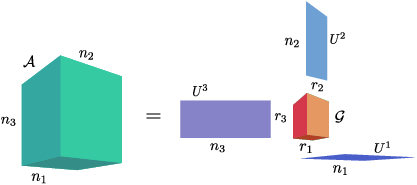}
\caption{Visual representation of Tucker decomposition.}
\label{fig:n1}
\end{center}
\end{figure} 

To compute the above\hyp mentioned Tucker components, one has to choose a suitable compression algorithm. The higher order singular value decomposition (HOSVD) is an orthogonal Tucker decomposition, widely used because it has a proven upper error bound \cite{Lathauwer2000}. Moreover, the algorithm is based entirely on singular value decomposition (SVD), which provides a robust and stable approximation of the initial tensor. Note that SVD requires the assembly of the initial array, which could be challenging for large tensors. In such cases, one could implement the Tucker decomposition using a 3D cross approximation algorithm \cite{Oseledets2008}. 

\subsubsection{Cross Approximation}

A matrix $A \in \mathbb{C}^{n_1 \times n_2}$ can be approximated with the so\hyp called 2D cross approximation method \cite{Tyrtyshnikov1996a, Tyrtyshnikov1996b} as follows:
\begin{equation}
A \approx U V^*.
\end{equation}
Here, $U \in \mathbb{C}^{n_1 \times r_c}$ and $V\in \mathbb{C}^{n_2 \times r_c}$. $r_c$ represents the column rank of matrix $A$. Cross approximation algorithms construct the decomposition of $A$ by using only some rows and columns of it, differently than SVD, which depends on the availability of the full matrix. Several algorithms have been developed over the previous decades for the implementation of cross approximation, with two being the most used ones: the ACA \cite{Kurz2002, Bebendorf2003}, and the maximum volume\hyp based cross algorithm \cite{Goreinov1997, Goreinov2001}. The latter requires the implementation of LU, QR and SVD for its efficient implementation. Therefore, the memory demand of the algorithm increases drastically for large tall matrices, such as the coupling matrices $T_{\rm bc}$ and $Z_{\rm bc}$ in the case fine voxel resolutions. On the other hand, the memory demand of ACA is only dictated by the size of matrices $U$ and $V$.  

\section{Tucker-Based Compression Algorithm}  \label{sc:III}

In VSIE, the columns of the coupling matrix describe interactions between coil and body basis functions through the Green's functions of equations (\ref{eq:nNt}) and (\ref{eq:nKt}); therefore, they represent well\hyp separated geometrical blocks. Due to the 3D nature of the problem, the key idea of our proposed compression algorithm is to reshape these columns as tensors and approximate them with the low multilinear Tucker model. This compression strategy enabled us to develop a new method to efficiently perform the matrix\hyp vector product and an extension to ACA, which are described later in this section.

\subsection{Matrix Assembly}

Each of the $m$ columns of the coupling matrices $Z_{\rm bc}$ and $T_{\rm bc}$ can be seen as the concatenation of $q$ vectors, where each vector represents the component\hyp wise interaction between one RWG element on the coil and the basis functions of all the voxels in the body domain. For PWC, $q = 3$, whereas for PWL, $q = 12$. Since these vectors model interactions between remote discretization elements, they have low\hyp rank properties \cite{Chai2013}. To exploit the low\hyp rank, each column of the coupling matrix can be reshaped as $q$ 3D tensors $\Zop^{kj} \in \mathbb{C}^{n_1\times n_2 \times n_3}$, $k = 1:q$, $n_{\rm v} = n_1\times n_2 \times n_3$, which are compressible with the Tucker decomposition \cite{francavilla2020maxwell}. A graphical description of the algorithm is shown in Fig. 2 for $Z_{\rm bc}$ and PWC basis functions,.   
    
\renewcommand{\thefigure}{2}
\begin{figure}[ht!]
\begin{center}
\includegraphics[width=0.48\textwidth, trim={0 0 0 0}]{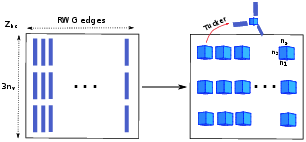}
\caption{Visual representation of the Tucker\hyp based algorithm for the compression of the $Z_{\rm bc}$ matrix, in the case of PWC basis functions. Each vector can be reshaped into a 3D tensor that is then compressed via Tucker decomposition.}
\label{fig:n2}
\end{center}
\end{figure} 

If the coupling matrix is first approximated with the ACA as $UV^*$, then our approach can still be used to compress the $r_c$ columns of $U$. In fact, cross approximation is a well\hyp conditioned operation, therefore the Tucker ranks of the reshaped columns of $U$ will be similar to the ones of the reshaped columns of the coupling matrix. The $V$ matrix here is usually much smaller than $U$ and does not require any additional compression.  
\par
TABLE II shows the total memory footprint associated with the assembly of the coupling matrix: Full assembly, assembly with ACA, and assembly with our proposed method by compressing either the columns of the coupling matrix (Tucker) or the columns of $U$ (ACA+Tucker). The memory required after compressing the coupling matrix with Tucker is $n_{\rm v} / \left(r^3 + 3nr\right)$ times smaller than the memory required by the full matrix, where $n$ and $r$ refer to the tensor's linear dimension and Tucker rank, respectively. If our Tucker\hyp based compression method is instead applied after the ACA assembly, then the total compression improves by a factor of $\sim m/r_c$, given that $r_c$ is small. TABLE II also shows the computational complexity of the assembly operations. The multiplicative constant factor $c_1$, which is present in all cases, represents the cost to compute the elements of the coupling matrix and is usually large. In fact, each element requires a 5D integration, whose computational cost depends on the number of both surface and volume quadrature integration points. As a result, the assembly of the matrix is extremely inefficient and should be implemented in parallel for multiple voxel\hyp RWG basis function interactions. 
\par 
Note that in certain cases, for example when the coil is close to the body, ACA may not achieve a meaningful compression and would not be advantageous to combine it with Tucker decomposition. In such cases, the preferable approach would be to divide the coupling matrix in $q$ blocks of size $n_{\rm v} \times m$, assembled them in parallel, and then compress their tensor components with a Tucker\hyp based method like HOSVD. Alternatively, if the coupling matrix is sufficiently small, one could assemble it in its full form and then apply Tucker directly to compress it.

\begin{table}[!ht]
\caption{Complexity for Constructing the Coupling Matrix} \label{tb:n2} \centering
{\def\arraystretch{2}\tabcolsep=3pt

\begin{tabular}{ c|c|c }
\hline
\hline
Assembly Method  & Operations                                                  & Memory                                   \\
\hline
Full             & $\Oop\left( c_1 q n_{\rm v} m \right)$                      & $q n_{\rm v} m   $                       \\    
ACA              & $\Oop\left( c_1 r_c^2 \left(qn_{\rm v} + m\right)  \right)$ & $q n_{\rm v} r_c + mr_c$                 \\   
Tucker           & Full + $\Oop\left( r q n_{\rm v} m \right)$                 & $q \left( r^3 + 3nr \right) m$           \\   
ACA+Tucker       & ACA + $\Oop\left( r q n_{\rm v} r_c \right)$                & $q \left( r^3 + 3nr \right) r_c  + mr_c$ \\   
\hline
\hline
\end{tabular}
}
\end{table}   

\subsection{Matrix-Vector Product}

Decompressing the coupling matrix to compute the matrix\hyp vector product $\matvec{y} = Z_{\rm bc} \matvec{x}$, like in equations (\ref{eq:n3}), may not be possible due to computer or GPU memory limitations. To address this, we propose a novel approach to efficiently compute the matrix\hyp vector product without fully decompressing the coupling matrix. We initiate $\matvec{y}$ as a vector of zeros. Inside a loop that cycles over the RWG basis functions, we decompress the $q$ tensors of a column $j \in [1,m]$, reshape them as vectors, and concatenate them to form the $j$\hyp th column of the original $Z_{\rm bc}$ matrix. The vector\hyp scalar product between the $j$\hyp th column and $\matvec{x}_j$ is then computed, and the result increments the elements of $\matvec{y}$. The same algorithm can be followed for the matrix\hyp matrix product $Y = Z_{\rm bc}X$.
\par
The conjugate transpose matrix\hyp vector product $\matvec{y} = Z^*_{\rm bc} \matvec{x}$ is required for the computation of the gradient of the cost function in the VSIE\hyp based GMT implementation. This case is slightly different than the standard matrix\hyp vector product: inside the loop cycling through the RWG functions, a row\hyp column vector product must be computed between the conjugate transpose of the decompressed $j$\hyp th column of $Z_{\rm bc}$ and $\matvec{x}$, which yields the scalar $\matvec{y}_j$. The algorithm remains instead the same for the conjugate transpose matrix\hyp matrix products. Both algorithms (for $p$\hyp columned matrices $X$ and $Y$) are summarized in the pseudocode below:

\begin{breakablealgorithm}
\caption{$\:Y = Z_{\rm bc} X$}\label{al:n1}
\begin{algorithmic}[1]
	\For{k=1:$q$}
	\State $Y^k = 0$
	\EndFor	
	\For{j=1:$m$} 
	\For{k=1:$q$}
	\State $\text{Decompress} \: \Zop^{kj}$
	\State $Y^k += \Zop^{kj}(:)\matvec{X_{j:}}$
	\EndFor
	\EndFor	
	\State $Y = \begin{bmatrix} Y^1 \\ \cdots \\ Y^q \end{bmatrix}$	
\end{algorithmic}
\end{breakablealgorithm}
\vspace{0.1cm}
\begin{breakablealgorithm}
\caption{$\:Y = Z^*_{\rm bc} X$}\label{al:n2}
\begin{algorithmic}[1]
	\State $Y = 0$
	\For{j=1:$m$} 
	\For{k=1:$q$}
	\State $\text{Decompress} \: \Zop^{kj}$
	\EndFor
	\State $Y_{j:} = \begin{bmatrix} \Zop^{1j}(:) \\ \cdots \\ \Zop^{qj}(:) \end{bmatrix}^* X$
	\EndFor	
\end{algorithmic}
\end{breakablealgorithm}

In Algorithm 1, $X \in \mathbb{C}^{m\times p}$, $Y \in \mathbb{C}^{qn_{\rm v}\times p}$ (vice\hyp versa for Algorithm 2), and $\Zop^{kj}(:)$ is the reshaped column vector of the tensor component $\Zop^{kj}$. The algorithms remain the same if $Z_{\rm bc}$ is compressed with ACA first ($Z_{\rm bc} = UV^*$). One has to replace $Z_{\rm bc}$ with $U$, $m$ with $r_c$, and assign $X = V^* X$ for Algorithm 1, and $Y = V Y$ for Algorithm 2. Both the standard and the conjugate transpose matrix\hyp vector products have the same complexity, shown in TABLE III, for the full, ACA, Tucker, and ACA+Tucker compressed cases. The full matrix\hyp vector product is faster than the Tucker\hyp compressed approach by a factor of $(r+p)/p$, which depends on the number of columns of $X$ and $Y$ and the Tucker rank. ACA can be faster than the full case for small values of $r_c$. Although the approach based on Tucker decomposition is slower because it requires additional flops compared to the other methods, it is more likely to fit in GPUs, thanks to its small memory footprint.  

\begin{table}[!ht]
\caption{Matrix\hyp Vector Product Complexity} \label{tb:n3} \centering
{\def\arraystretch{2}\tabcolsep=3pt

\begin{tabular}{ c|c }
\hline
\hline
Matrix Form       & Operations Complexity \\
\hline
Full              & $\Oop\left( q n_{\rm v} m p\right)$                                                                           \\    
ACA               & $\Oop\left( q n_{\rm v} r_c p\right)$  + $\Oop\left( r_c m p\right)$                                          \\
Tucker            & $\Oop\left( r q n_{\rm v} m \right)$   + $\Oop\left( q n_{\rm v} m p \right)$                                 \\
ACA+Tucker        & $\Oop\left( r q n_{\rm v} r_c \right)$ + $\Oop\left( q n_{\rm v} r_c p \right)$ + $\Oop\left( r_c m p\right)$ \\
\hline
\hline
\end{tabular}
}
\end{table} 
 
\subsection{Tucker\hyp based ACA}

If the coupling matrix is first compressed with ACA, the previous methods for matrix assembly and matrix\hyp vector product could still be applied to the matrix $U$ of the cross approximation. However, for the case of realistic body models discretized with fine voxel resolutions, the traditional implementation of ACA (a detailed description can be found in \cite{Zhao2005}) might fail due to random access memory (RAM) overflow because of the size of $U$ (see section IV.B.2). To address this, we propose an extension of ACA in which the matrix $U$ is assembled directly in a compressed form, based on our proposed Tucker decomposition technique. The algorithm is summarized in pseudocode bellow:  

\begin{breakablealgorithm}
\caption{ACA of $Z\in \mathbb{C}^{m_1 \times m_2}$. Assembly with compressed $U$}\label{al:n3}
\begin{algorithmic}[1]

	\State $\epsilon = \text{tolerance}$, $i = 1$, $\matvec{s}_1 = 0$
	\State $\mathscr{U} = []$, $V = []$

	\For{$k=1:\text{min}(m_1,m_2)$}

	\State $\matvec{r} \gets Z_{i:}$

	\If {$k > 1$}

	\State $[f_1,f_2,f_3,p] \gets i$

	\For{$l=1:size(\mathscr{U},2)$}
	\State $[\Gop,U^1,U^2,U^3] \gets U_{pl}$
	\State $\matvec{t}_l = \Gop \times_1 U_{f_1,:}^1 \times_2 U_{f_2,:}^2 \times_3 U_{f_3,:}^3$
	\EndFor
		
	\State $\matvec{r} \pluseq - \matvec{t}V^*$
	\EndIf

	\State $j \gets \text{index of max element of}\: \matvec{r}$
	\State $\matvec{y} \gets \left( \matvec{r}/\matvec{r}_j \right)^*$
	\State $\matvec{x} \gets Z_{:j}$

	\If {$k > 1$}
	\State $\matvec{x} \pluseq - \textbf{Alg1}(\mathscr{U},V_{j:}^*)$
	\EndIf
	
	\State $\matvec{s}_{k+1} \gets \matvec{s}_k + (\norm{\matvec{x}}\norm{\matvec{y}})^2$	
	
	\If {$k > 1$}
	\State $\matvec{s}_{k+1} \pluseq 2\sum\left[\Re\{\left(\textbf{Alg2}(\mathscr{U},\matvec{x})\right) \odot \left(V^*\matvec{y} \right)\}\right]$
	\EndIf
	
	\State Reshape $\matvec{x}$ to $q$ $\Xop^q$ tensors.
	\State $\mathscr{U} = [\mathscr{U} \: \text{HOSVD}(\Xop^1, 3 \epsilon) \: \cdots \: \text{HOSVD}(\Xop^q, 3 \epsilon)]$
	\State $V = [V \: y]$
	
	\IfThen {$\norm{\matvec{x}}\norm{\matvec{y}} \leq \epsilon \sqrt{\matvec{s}_{k+1}}$} {\textit{break}}
	
	\State $\matvec{x} = \abs{\matvec{x}}$, \: $\matvec{x}_i = 0$
	\State $i \gets \text{index of max element of}\: \matvec{x}$
		
	\EndFor	
\end{algorithmic}
\end{breakablealgorithm}   

Here $\mathscr{U}$ is a struct of size $q \times r_c$ ($q=3$ for PWC, $q=12$ for PWL, $r_c$ is the rank of $Z$) than contains tensors. Each time a new column of $U$ is computed, it is reshaped to $q$ tensors, which are then compressed with a truncated HOSVD of tolerance $3 \epsilon$ (line 19,20). The HOSVD tolerance has to be higher than the ACA tolerance since the irrelevant numerical digits ($<1e-3$) appearing in $U$ are incompressible. We found that a $3$ times higher tolerance is a good choice for our numerical examples. To perform matrix\hyp \ and conjugate transpose matrix\hyp vector products with the compressed $U$ we followed Algorithms 1 (line 15) and 2 (line 18). Finally, when a row of $U$ is requested in Algorithm 3, we first calculate the voxel and basis function component corresponding to that row (line 6) and then decompress, using equation (6), only the required elements from the Tucker compressed components of $U$ (line 8,9). The proposed algorithm avoids RAM overflowing, but it is slower than the traditional ACA due to the multiple tensor decompressions. Nevertheless, it could always be accelerated via GPU, since its memory demand is as low as for one column of the coupling matrix.
         
\section{Numerical Experiments} \label{sc:IV}

\subsection{Tucker Rank Behavior} \label{ssc:A}

In this section, we study the low\hyp Tucker rank properties of the $Z_{\rm bc}$ coupling matrix. We considered multiple geometrical scenarios and altered the distance between the conductive surface (coil) and the VIE domain, the operating frequency and the conductive surface's discretization. The tensor components of the columns of the coupling matrix were compressed with the HOSVD algorithm and a tolerance of $1e-8$, which yielded a relative error similar to the tolerance for all cases, due to the robustness of the SVD itself. Such error can be considered negligible, because the tolerance of the iterative solver used in FFT\hyp based VIE systems is usually orders of magnitude higher. 

\subsubsection{Tucker Rank vs. Distance} \label{ssc:A1}

It is well established that the Green's function integro\hyp differential operators between well\hyp separated geometries present low\hyp rank properties \cite{Chai2013}. Here we studied the relation between the low multilinear rank of the compressed coupling matrix $Z_{\rm bc}$ and the distance between the body's domain and the coil. We set the frequency to $298.06$ MHz, the operating frequency of $7$ Tesla MRI. We modeled a single perfectly electric conducting (PEC) loop coil of radius $\rho = 0.50$ m and discretized it with $125$ triangular elements. The coil was centered in $\left(0,d,0\right)$, where $d$ was varied as $0.55$, $0.6$, $\dots$, $1$ m. The domain was a cuboid with edge length of $1$ m, centered at $(0,0,0)$ and discretized with voxels of $1$ cm isotropic resolution and PWC basis functions (Fig. 3). As a result, the tensor's size was $101\times 101\times 101$ and the memory required by the fully assembled $Z_{\rm bc}$ was $5848$ MBs. 

\renewcommand{\thefigure}{3}
\begin{figure}[ht!]
\begin{center}
\includegraphics[width=0.48\textwidth, trim={0 0 0 0}]{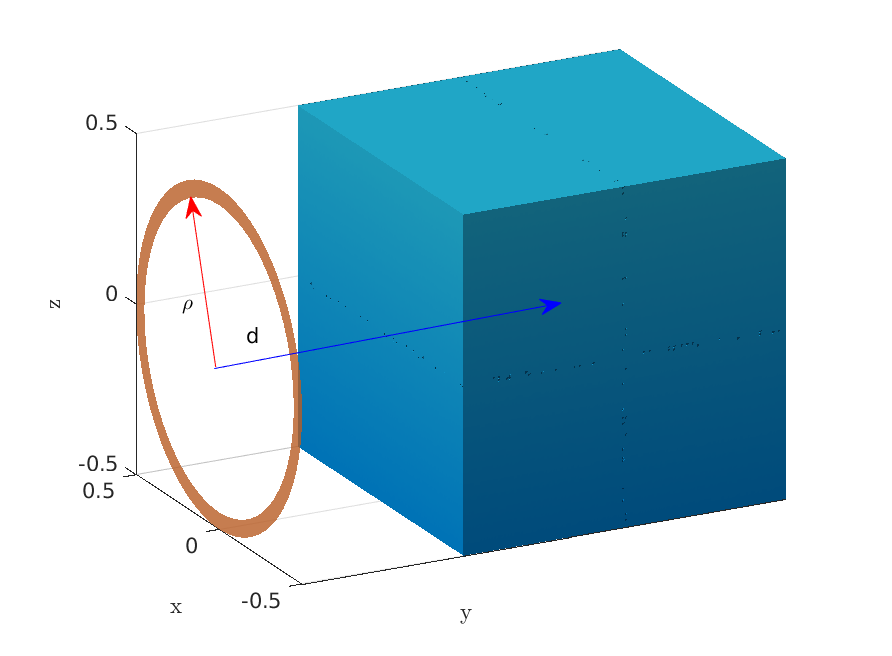}
\caption{Loop\hyp cubic domain geometry. The loop coil was shifted on the $\hat{y}$ direction, for $10$ discrete distances between $0.55$ to $1$ m from the center of the cube.}
\label{fig:n3}
\end{center}
\end{figure} 

Fig. 4 illustrates the reduction of the maximum rank (maximum of all Tucker ranks for all components) of the coupling matrix (right axis), and the total memory of the compressed matrix using the algorithm described in Section IV (left axis). It is evident that the greater the distance between the domain and the coil, the lower the multilinear ranks and the required memory. The compression factor varied between $\sim 50$ and $190$, depending on the distance. 

\renewcommand{\thefigure}{4}
\begin{figure}[ht!]
\begin{center}
\includegraphics[width=0.48\textwidth, trim={0 0 0 0}]{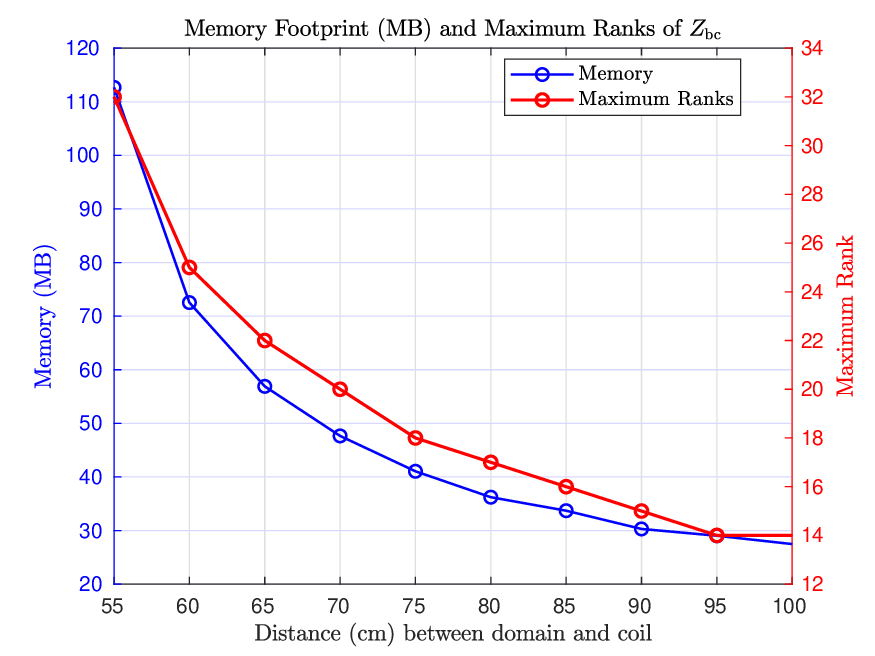}
\caption{Memory footprint (left) and maximum rank (rank) of the compressed $Z_{\rm bc}$ matrix, for different distances between the loop and the cubic domain.}
\label{fig:n4}
\end{center}
\end{figure} 

\subsubsection{Tucker Rank vs. Frequency} \label{ssc:A2}

The work presented in \cite{Chai2013} showed that the rank of the integral operators for 3D problems increases linearly with respect to the operating frequency. This was confirmed in \cite{giannakopoulos2019memory}, for the BTTB defining tensors of the FFT\hyp based VIE systems (discretized integral operators). These tensors were columns of the corresponding Galerkin MoM matrices and modeled the interactions between one voxel's basis function and the whole domain via the $\Nt$ or $\Kt$ operators. In the present study, the tensors are columns of the coupling matrix and model the interactions between one RWG basis function and the whole body domain via the same operators. Since in both cases the interactions between separated geometry blocks are modeled, one can expect a similar behavior for the Tucker ranks.
\par  
To confirm this, we performed a frequency sweep ($300$, $600$, $\dots$, $2700$MHz) for the setup in Fig. 3. The coil was discretized with $125$ elements, whereas the voxel's isotropic resolution was set to $\lambda / 20$, with $\lambda$ being the wavelength. We repeated the calculations for three positions of the coil ($d=0.55$, $0.65$, and $0.8$ m). The memory footprint (left) of the compressed matrix, along with the maximum rank (right), are shown in the dual axis chart of Fig. 5. The memory footprint increased linearly with frequency, whereas the maximum rank grew at different rates for the three investigated cases. This is expected because the maximum rank represents the worst\hyp case scenario among all ranks, whereas the memory footprint summarizes the overall effect of all ranks. 

\renewcommand{\thefigure}{5}
\begin{figure}[ht!]
\begin{center}
\includegraphics[width=0.48\textwidth, trim={0 0 0 0}]{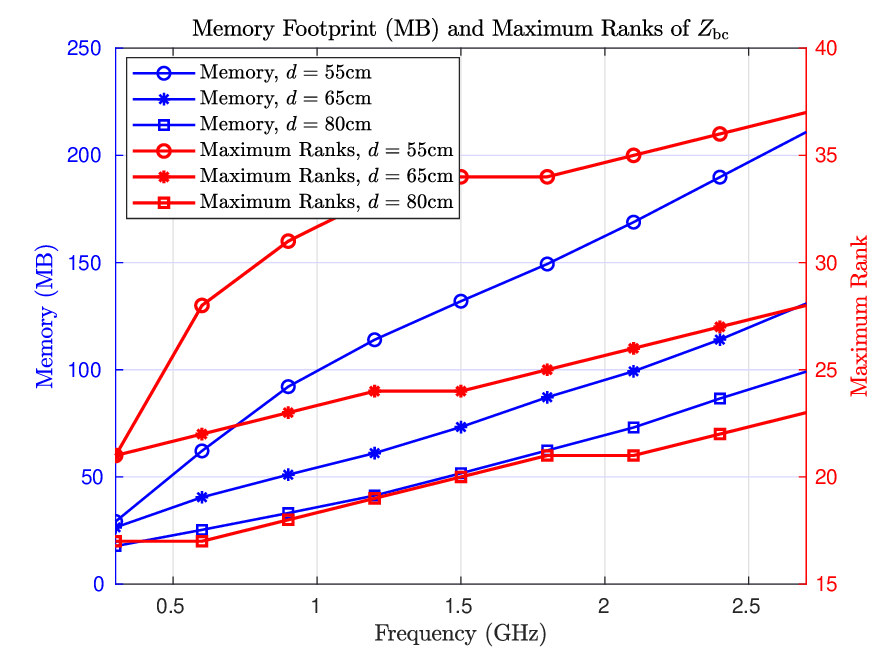}
\caption{Memory footprint (left) and maximum rank (rank) of the compressed $Z_{\rm bc}$ matrix, for different operating frequencies. Results are shown for three different distances between the loop and the domain.}
\label{fig:n5}
\end{center}
\end{figure} 

\subsubsection{Tucker Rank vs. Surface Mesh} \label{ssc:A3}

Let us consider a fixed mesh for the domain and refine only the surface mesh of the coil. As the coil mesh is refined, the edges of the triangular elements become smaller and the number of columns of the coupling matrix increases, with each column representing more remote element interactions between an edge of the triangular mesh and the voxelized domain. As a result, we should expect a drop in the Tucker ranks. To verify this, we used the same domain of the previous sections, and a PEC equilateral triangle with centroid at $(0,0.55\text{m},0)$ and one vertex at $(0,0.55\text{m},0.5\text{m})$. The triangle was discretized with $10$ different meshes, starting from triangular element's edge of $0.5$m and reducing it by a factor of $\sqrt{2}$, which resulted in $4$, $6$, $11$, $30$, $48$, $102$, $184$, $358$, $727$, and $1480$ elements. Fig. 6 reports the maximum rank as a function of the length of the triangular element's edge, confirming that the rank is smaller when the PEC triangle's mesh is finer.

\renewcommand{\thefigure}{6}
\begin{figure}[ht!]
\begin{center}
\includegraphics[width=0.48\textwidth, trim={0 0 0 0}]{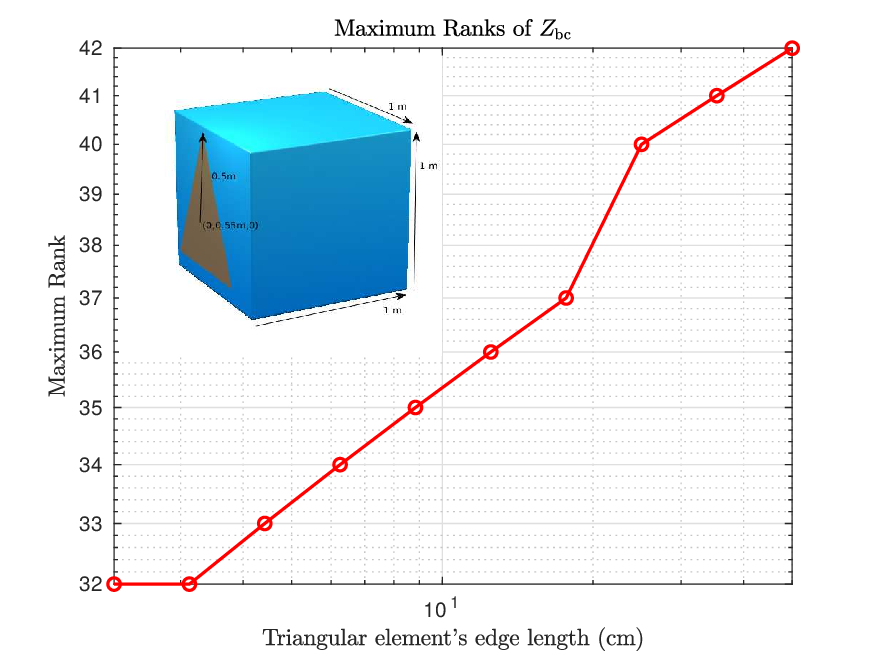}
\caption{Maximum rank of the compressed $Z_{\rm bc}$ matrix, for various PEC triangle's meshes. The rank drops as we refine the mesh.}
\label{fig:n6}
\end{center}
\end{figure} 

\subsection{Application to VSIE-based MRI Simulations}

Here we aim to validate the performance of the proposed algorithms for the assembly of the coupling matrix $Z_{\rm bc}$, and the matrix\hyp vector implementation for two VSIE\hyp based MRI applications. Both numerical experiments were implemented in Matlab, except for the matrix assembly part which was written in \CC. For the GPU computations, we used an NVIDIA Quadro Volta GV100 32GB HBM2 PCIe. For the CPU computations, in the first experiment we used a server with CentOS 6.9 operating system and an Intel(R) Xeon(R) CPU E5\hyp 2699 v3 at 2.30GHz, while for the second experiment we used a server with Ubuntu 18.04.5 LTS operating system and an Intel(R) Xeon(R) Gold 6248 CPU at 2.50GHz. We parallelized on $12$ workers where needed. 

\subsubsection{Head Coil Experiments}

We first demonstrated the proposed compression method for an 8-ports close-fitting head coil, previously designed for GMT \cite{giannakopoulos2020magnetic}, which we loaded with the ``Billie'' realistic head model from the virtual family population \cite{VirtualFamily} (Fig. 7). The operating frequency was set to $298$ MHz.

\renewcommand{\thefigure}{7}
\begin{figure}[ht!]
\begin{center}
\includegraphics[width=0.48\textwidth, trim={0 0 0 0}]{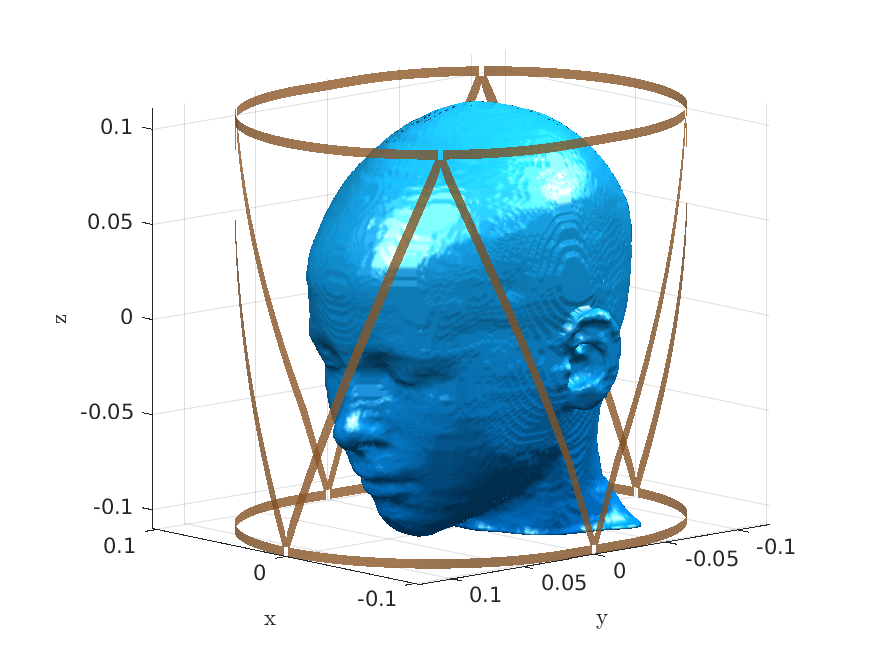}
\caption{Coil\hyp head geometry. The RF coil (discretized with $2380$ triangular element edges) was loaded with the voxelized realistic human head model ``Billie'' (discretized with voxels of $1$ mm isotropic resolution).}
\label{fig:n7}
\end{center}
\end{figure} 

The VSIE-based implementation of GMT requires performing operations on the coupling matrix $Z_{\rm bc}$ and its conjugate transpose. We analyzed the memory footprint reduction for the compressed coupling matrix and measured the computation time for both the matrix\hyp \ and conjugate transpose matrix\hyp vector products using the algorithms presented in section III. The coil was discretized with both a coarse ($516$ RWG) and a fine ($2380$ RWG) mesh resolution. For the VIE domain enclosing the head, we tested three different voxel resolutions, namely $5$, $2$, and $1$ mm$^3$, which resulted in $34 \times 38 \times 45$, $84 \times 96 \times 116$, and $168 \times 188 \times 222$ voxels, respectively. Both PWC ($3$ unknowns per voxel) and PWL ($12$ unknown per voxel) VIE basis functions were considered. 
\par
We used a tolerance of $1e-8$ for HOSVD, which would ensure accurate estimation of electrical properties in an actual GMT experiment. Since the coil closely fits the head, ACA (or SVD\hyp based methods in general) are expected to provide negligible compression with a tolerance of $1e-8$. We confirmed this for the case of PWC basis functions, $5$ mm$^3$ voxel resolution, and fine coil discretization, for which, in fact, we found that $2238$ of the $2380$ singular values would be needed to accurately represent $Z_{\rm bc}$, compressing the matrix from $6.18$ GB to $6.07$ GB. Consequently, for the head coil experiments we did not use the Tucker\hyp based ACA algorithm, but instead we compressed the columns of the coupling matrix only with the HOSVD\hyp based method.

\paragraph{Memory Compression}

The memory footprint for the assembly of the coupling matrix $Z_{\rm bc}$ is shown in TABLE IV. The memory required to assemble the full matrix was considerably larger than for the HOSVD-compressed matrix. For example, for PWC basis functions, voxel resolution of $1$ mm$^3$, and fine coil mesh, the required memory in the full matrix case was $>740$ GBs, whereas the compressed matrix required only $2.6$ GBs. Note that in the challenging case of PWL basis functions, $1$ mm$^3$ voxel resolution, and fine coil mesh, it was not feasible to apply our compression method. In fact, the memory requirements even for just one of the $q$ blocks of the matrix (see Section IV. A.) were prohibitively large for our server. While we could have still implemented the HOSVD compression by further dividing the matrix in smaller blocks, that would have required $\sim 1$ month of computations. An alternative method for such costly cases is mentioned in the discussion and will be pursued in future work. 

\begin{table}[!ht]
\caption{Memory Requirements (GBs) of $Z_{\rm bc}$} \label{tb:memory} \centering
{\def\arraystretch{2}\tabcolsep=2pt

\begin{tabular}{ c|c|c|c|c|c }
\hline
\hline
Voxel Res.                  & Assembly & PWC\hyp coarse & PWC\hyp fine   & PWL\hyp coarse & PWL\hyp fine \\
\hline
\multirow{2}{*}{$5$ mm$^3$} & Full     & 1.34           & 6.18           & 5.36           & 24.74    \\
						    & HOSVD    & 0.20           & 0.88           & 0.86           & 3.77     \\
\hline
\multirow{2}{*}{$2$ mm$^3$} & Full     & 21.57          & 99.52          & 86.30          & 398.09   \\
						    & HOSVD    & 0.40           & 1.74           & 1.79           & 7.49     \\
\hline
\multirow{2}{*}{$1$ mm$^3$} & Full     & 161.73         & 745.99         & 646.95         & 2983.99  \\
						    & HOSVD    & 0.63           & 2.62           & 2.85           & N/A  \\
\hline
\hline
\end{tabular}
}
\end{table}

Fig. 8 shows that the compression factor, defined as the memory of the full matrix over the memory of the compressed one, decreased as the voxel resolutions ($h$) of the VIE domain's grid became coarser. The behavior of the compression factor was similar for PWC or PWL basis functions, either with fine or coarse mesh discretization. This confirms the excellent stability of our compression method.

\renewcommand{\thefigure}{8}
\begin{figure}[ht!]
\begin{center}
\centering
\includegraphics[width=0.48\textwidth, trim={0 0 0 0}]{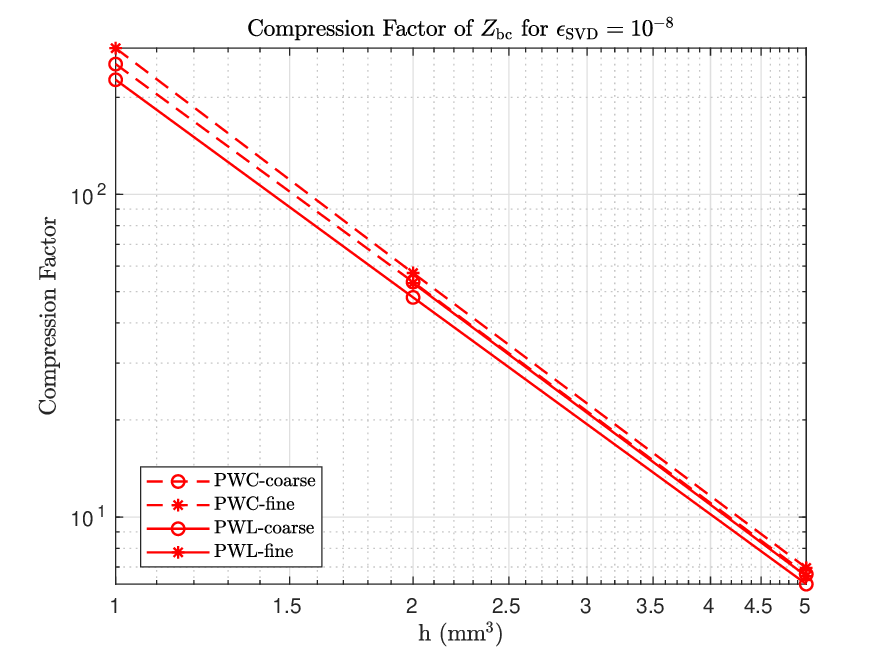}	
\caption{Compression factor of the compressed matrix $Z_{\rm bc}$. Results are shown for all investigated head and coil discretizations.}
\label{fig: n8}
\end{center}
\end{figure}

Fig. 9 shows the maximum Tucker rank, obtained with HOSVD, for all tensor components of the coupling matrix. The rank decreased slowlier than the compression factor (Fig. 8) for coarser discretizations of the VIE domain. For example, when PWC basis functions and fine coil resolution were used (PWC-fine), the maximum rank decreased by only $1.5$ times (from $42$ to $28$) when the isotropic voxel size increased from $1$ to $5$ mm$^3$, which corresponds to a $5$ times smaller grid in all directions. For all cases, the maximum rank was smaller for finer coil meshes, which is in agreement with the results shown in section V.A.3. 

\renewcommand{\thefigure}{9}
\begin{figure}[ht!]
\begin{center}
\centering
\includegraphics[width=0.48\textwidth, trim={0 0 0 0}]{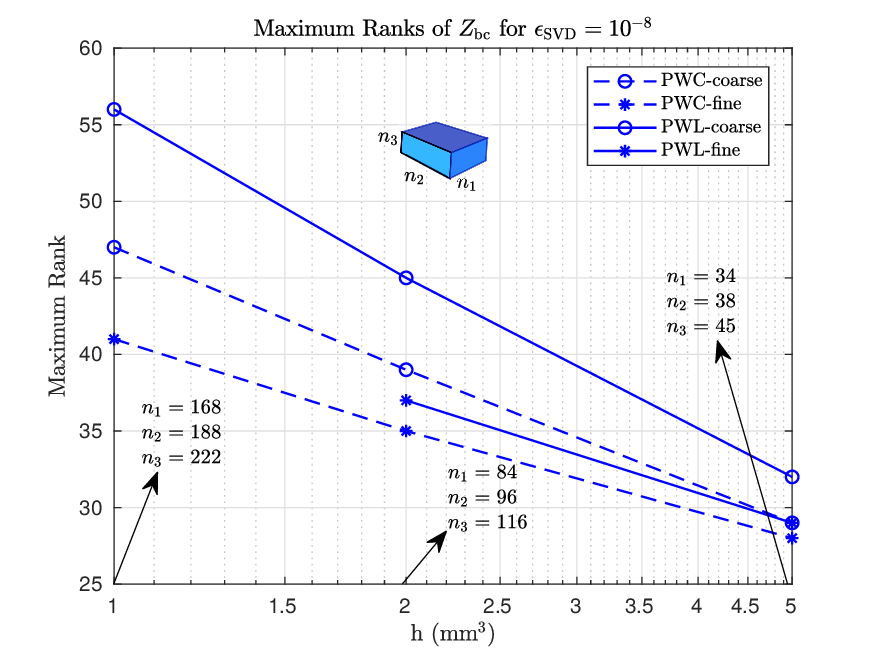}	
\caption{Maximum rank as a function of voxel resolution of the VIE domain. The maximum rank was calculated among all Tucker ranks of the decomposed tensors, which were obtained with HOSVD. Results are shown for all investigated head and coil discretizations. For each voxel resolution, the size of the corresponding VIE discretization grid is indicated.}
\label{fig: n9}
\end{center}
\end{figure}

\paragraph{Computation Time}

TABLE V reports the computation time for the assembly of the full and the HOSVD\hyp compressed coupling matrix (rounded to the nearest higher second). For PWC, we used one quadrature integration point per triangle and voxel, while for PWL, two for each triangle and eight for each voxel. For the low resolution matrices, the assembly time for the compressed matrix was larger than the one for the full matrix by a factor $\leq q$, since in the HOSVD case the compressed matrix was assembled as $q$ sequential compressed matrix blocks. For the larger matrices, our server could not perform the assembly of the full matrix, due to the prohibitively large memory requirements, but it was able to assemble the compressed matrix using our proposed method.

\begin{table}[!ht]
\caption{Time Footprint (hh:mm:ss) of $Z_{\rm bc}$ Assembly} \label{tb:time_assembly} \centering
{\def\arraystretch{2}\tabcolsep=2pt

\begin{tabular}{ c|c|c|c|c|c }
\hline
\hline
Voxel Res.                  & Assembly & PWC\hyp coarse & PWC\hyp fine   & PWL\hyp coarse & PWL\hyp fine \\
\hline
\multirow{2}{*}{$5$ mm$^3$} & Full     & 00:00:12       & 00:00:45       & 00:03:31       & 00:14:17 \\
						    & HOSVD    & 00:00:34       & 00:02:33       & 00:32:27       & 02:21:21 \\
\hline
\multirow{2}{*}{$2$ mm$^3$} & Full     & 00:02:48       & N/A            & 01:09:49       & N/A      \\
						    & HOSVD    & 00:06:16       & 00:33:28       & 07:15:45       & 30:32:42 \\
\hline
\multirow{2}{*}{$1$ mm$^3$} & Full     & N/A            & N/A            & N/A            & N/A      \\
						    & HOSVD    & 00:27:59       & 02:31:52       & 52:17:11       & N/A      \\
\hline
\hline
\end{tabular}
}
\end{table}
     
TABLE VI and VII summarize the computation times for the matrix\hyp \ and conjugate transpose matrix\hyp vector products. Compared to the full form case (Full\hyp CPU), the compressed matrix\hyp vector product requires additional operations for the decompression of the tensors. While Algorithms 1 and 2 can reduce the memory requirements of the matrix\hyp vector products, the time footprint varies based on how these algorithms are implemented. In particular, the nested loops over the $m$ RWG functions can be either parallelized on a CPU, if the RAM can support multiple tensor decompressions in parallel (HOSVD-CPU), or performed sequentially using a GPU (HOSVD-GPU). In our tests, we multiplied $Z_{\rm bc}$ with $X\in \mathbb{C}^{m \times 8}$ (TABLE VI) and $Z^*_{\rm bc}$ with $\Phi \in \mathbb{C}^{qn_{\rm v} \times 8}$ (TABLE VII), where both $X$ and $\Phi$ were random matrices to keep the results general. The eight columns of $X$ could correspond, for example, to the currents associated with the eight channels of the coil in Fig. 7.

\begin{table}[!ht]
\caption{Time Footprint (hh:mm:ss) of $Y = Z_{\rm bc} X$} \label{tb:time_forward_matvec} \centering
{\def\arraystretch{2}\tabcolsep=2pt

\begin{tabular}{ c|c|c|c|c|c }
\hline
\hline
Voxel Res.                  & Form          & PWC\hyp coarse & PWC\hyp fine   & PWL\hyp coarse & PWL\hyp fine \\
\hline
\multirow{3}{*}{$5$ mm$^3$} & Full\hyp CPU  & 00:00:01       & 00:00:02       & 00:00:02       & 00:00:06 \\
						    & HOSVD\hyp CPU & 00:00:03       & 00:00:07       & 00:00:07       & 00:00:24 \\
						    & HOSVD\hyp GPU & 00:00:03       & 00:00:11       & 00:00:10       & 00:00:44 \\
\hline
\multirow{3}{*}{$2$ mm$^3$} & Full\hyp CPU  & 00:00:06       & N/A            & 00:00:25       & N/A      \\
						    & HOSVD\hyp CPU & 00:00:25       & 00:01:46       & 00:01:37       & 00:06:24 \\
						    & HOSVD\hyp GPU & 00:00:04       & 00:00:14       & 00:00:13       & 00:00:56 \\
\hline
\multirow{3}{*}{$1$ mm$^3$} & Full\hyp CPU  & N/A            & N/A            & N/A            & N/A      \\
						    & HOSVD\hyp CPU & 00:02:54       & 00:11:25       & 00:11:30       & N/A      \\
						    & HOSVD\hyp GPU & 00:00:13       & 00:00:53       & 00:00:52       & N/A      \\
\hline
\hline
\end{tabular}
}
\end{table}

\begin{table}[!ht]
\caption{Time Footprint (hh:mm:ss) of $\Psi = Z^*_{\rm bc} \Phi$} \label{tb:time_adjoint_matvec} \centering
{\def\arraystretch{2}\tabcolsep=2pt

\begin{tabular}{ c|c|c|c|c|c }
\hline
\hline
Voxel Res.                  & Form          & PWC\hyp coarse & PWC\hyp fine & PWL\hyp coarse & PWL\hyp fine \\
\hline
\multirow{3}{*}{$5$ mm$^3$} & Full\hyp CPU  & 00:00:01       & 00:00:02     & 00:00:02       & 00:00:06 \\
						    & HOSVD\hyp CPU & 00:00:02       & 00:00:04     & 00:00:04       & 00:00:26 \\
						    & HOSVD\hyp GPU & 00:00:02       & 00:00:07     & 00:00:05       & 00:00:21 \\
\hline
\multirow{3}{*}{$2$ mm$^3$} & Full\hyp CPU  & 00:00:06       & N/A          & 00:00:22       & N/A      \\
						    & HOSVD\hyp CPU & 00:00:13       & 00:00:45     & 00:00:49       & 00:03:25 \\
						    & HOSVD\hyp GPU & 00:00:03       & 00:00:11     & 00:00:10       & 00:00:42 \\
\hline
\multirow{3}{*}{$1$ mm$^3$} & Full\hyp CPU  & N/A            & N/A          & N/A            & N/A      \\
						    & HOSVD\hyp CPU & 00:01:20       & 00:05:19     & 00:05:23       & N/A      \\
						    & HOSVD\hyp GPU & 00:00:11       & 00:00:45     & 00:00:41       & N/A      \\
\hline
\hline
\end{tabular}
}
\end{table}

For $5$ mm$^3$ isotropic voxel resolution, the Full\hyp CPU matrix\hyp vector product was the fastest for all cases, because the coupling matrix is small. For $2$ and $1$ mm$^3$ voxel resolution, the HOSVD\hyp GPU implementation was the fastest. Note that the Full\hyp CPU case could not be performed for high voxel and coil mesh resolutions, due to the excessive memory requirements. The HOSVD\hyp CPU was slower than HOSVD\hyp GPU, except for the $5$ mm$^3$ voxel resolution.

\subsubsection{Body Coil Experiments}

For the second MRI experiment, we simulated the volume bodycoil of a commercial 3T MRI scanner \cite{siemens2017magnetom, milshteyn2021individualized} and we loaded it with ``Billie'', from the virtual family population \cite{VirtualFamily} (Fig. 10). The frequency was set to $123$ MHz, corresponding to $3$ Tesla MRI. The coil has $32$ legs, a radius of $35.5$ cm, length of $45$ cm, and is centered at $\left(0,0,0\right)$. We also modeled the system conductive shield, which has a radius of $37.2$ cm, a length of $1.5$ m and is centered at $\left(0,0,0\right)$. The distance between the coil and the cuboid VIE domain enclosing ``Billie'' was $15.5$ cm and $24.5$ cm in the $x$ and $y$, respectively. In contrast with the previous case where the coil tightly fitted the head, here the coil is remote enough to allow a good compression of the coupling matrix $Z_{\rm bc}$ with ACA. For this experiment, we used PWC and PWL basis functions and three voxel resolutions ($5$, $2$, and $1$ mm$^3$), which corresponded to $81 \times 44 \times 108$, $205 \times 109 \times 270$, and $409 \times 219 \times 541$ voxels for the VIE domain. For the coil and the shield we used $9450$ RWG basis functions. Two quadrature integration points were used for each triangle and eight for each voxel, both for PWC and PWL basis functions. 

\renewcommand{\thefigure}{10}
\begin{figure}[ht!]
\begin{center}
\includegraphics[width=0.48\textwidth, trim={0 0 0 0}]{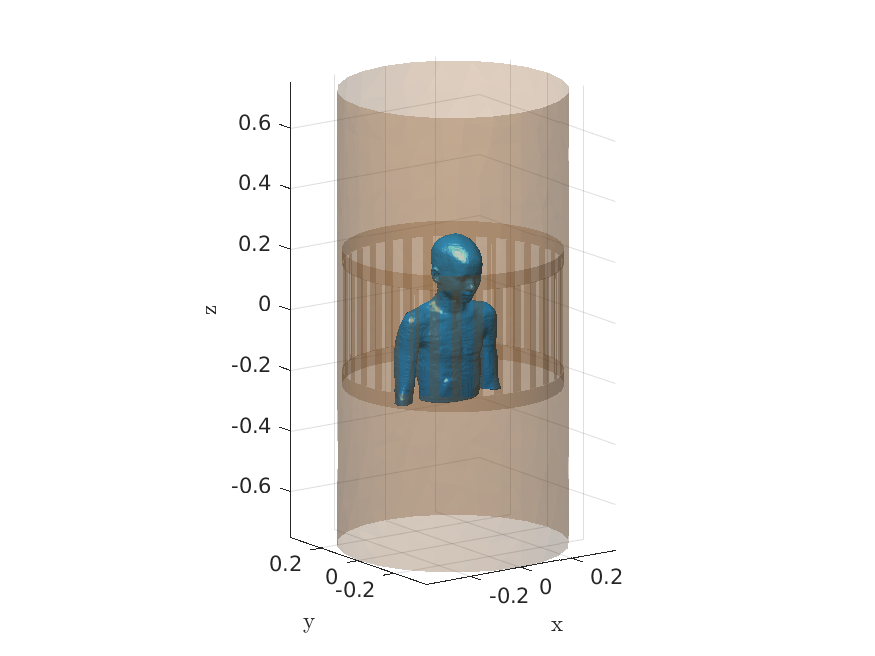}
\caption{Coil\hyp body geometry. The RF coil and shield (discretized with $9450$ triangular element edges) was loaded with part of the voxelized realistic human body model ``Billie'' (discretized with voxels of $2$ mm isotropic resolution).}
\label{fig:n10}
\end{center}
\end{figure} 

\paragraph{Matrix Assembly}

TABLE VIII, summarizes the memory requirements and the assembly time for the coupling matrix $Z_{\rm bc}$. The ACA tolerance was set to $1e-3$ to achieve good compression. The ACA rank of $Z_{\rm bc}$ was $250$ for the $5$ mm$^3$ cases and $287$ for the $2$ and $1$ mm$^3$ cases. The maximum Tucker rank of $U$ was between $15$ and $18$ for all cases. In the $5$ mm$^3$ case, our results show that ACA could offer an excellent compression of the coupling matrix and the assembly could be rapidly performed in CPU. For $2$ mm$^3$, ACA's memory requirements were large and ACA was outperformed in speed by our proposed Algorithm 3 (Tucker\hyp based ACA), for which the low memory footprint allowed using a GPU. For $1$ mm$^3$ resolution, the standard ACA algorithm could not be performed even on a server equipped with hundreds of GB's of RAM, due to overwhelming memory requirements. On the other hand, our proposed ACA extension in Algorithm 3 kept the memory demand small, enabling for fast matrix assembly in GPU. Note that the full matrix assembly was only possible for $5$ mm$^3$ voxel resolution and PWC basis functions. 

\begin{table}[!ht]
\caption{Memory Demand (GB) and \protect\linebreak Time Footprint (hh:mm:ss) of $Z_{\rm bc}$ Assembly} \label{tb:assembly2} \centering
{\def\arraystretch{2}\tabcolsep=2pt

\begin{tabular}{ c|c|c|c|c|c }
\hline
\hline
\multirow{2}{*}{Voxel Res.} & \multirow{2}{*}{Form} & \multicolumn{2}{c}{PWC} & \multicolumn{2}{c}{PWL} \\
\cline{3-6}
                            &                       & memory & time           & memory & time           \\
\hline
\multirow{3}{*}{$5$ mm$^3$} & Full\hyp CPU          & 162.60 & 00:15:49       & 650.42 & N/A            \\
						    & ACA\hyp CPU           & 4.33   & 00:01:53       & 17.24  & 00:06:39       \\
						    & Algorithm 3\hyp GPU   & 0.036  & 00:03:26       & 0.036  & 00:09:58       \\
\hline
\multirow{3}{*}{$2$ mm$^3$} & Full\hyp CPU          & 2548   & N/A            & 10195  & N/A            \\
					        & ACA\hyp CPU           & 77.44  & 00:37:16       & 309.65 & 04:30:36       \\
						    & Algorithm 3\hyp GPU   & 0.041  & 00:26:57       & 0.042  & 01:28:53       \\
\hline
\multirow{3}{*}{$1$ mm$^3$} & Full\hyp CPU          & 20471  & N/A            & 81884  & N/A            \\
							& ACA\hyp CPU           & 621.75 & N/A            & 2486   & N/A            \\
						    & Algorithm 3\hyp GPU   & 0.041  & 03:35:36       & 0.042  & 15:20:41       \\
\hline
\hline
\end{tabular}
}
\end{table}

For the coarser case of $5$ mm$^3$ voxel resolution and PWC basis functions, the time footprint of Algorithm 3 for CPU (not shown in the table) was 00:17:50, which is $\sim 5$ times slower than for the GPU execution.

\paragraph{Matrix\hyp Vector Product Performance}

The time footprints for the matrix\hyp vector product between the compressed coupling matrix $Z_{\rm bc}$ and a random vector $\matvec{x} \in \mathbb{C}^{m\times 1}$ are shown in TABLE IX. ACA-CPU corresponds to performing the product $U(V^*\matvec{x})$ in CPU. For ACA+HOSVD\hyp GPU, $Z_{\rm bc}$ was compressed with Algorithm 3, and the matrix\hyp vector product was performed with Algorithm 1 in GPU. For $5$ mm$^3$ voxel resolution, the efficiency is similar for both approaches. For $2$ mm$^3$ voxel resolution, ACA+HOSVD\hyp GPU outperformed ACA by a factor of $3$, because, due to its low memory demand, it could be executed on a GPU, whereas ACA could not. 

\begin{table}[!ht]
\caption{Time Footprint (hh:mm:ss) of $\matvec{y} = Z_{\rm bc} \matvec{x}$} \label{tb:time_forward_matvec2} \centering
{\def\arraystretch{2}\tabcolsep=2pt

\begin{tabular}{ c|c|c|c }
\hline
\hline
Voxel Res.                  & Form              & PWC       & PWL        \\
\hline
\multirow{3}{*}{$5$ mm$^3$} & Full\hyp CPU      & 00:00:11  & N/A        \\
						    & ACA\hyp CPU       & 00:00:01  & 00:00:01   \\
						    & ACA+HOSVD\hyp GPU & 00:00:01  & 00:00:02   \\
\hline
\multirow{3}{*}{$2$ mm$^3$} & Full\hyp CPU      & N/A       & N/A        \\
					        & ACA\hyp CPU       & 00:00:05  & 00:00:18   \\
						    & ACA+HOSVD\hyp GPU & 00:00:02  & 00:00:06   \\
\hline
\multirow{3}{*}{$1$ mm$^3$} & Full\hyp CPU      & N/A       & N/A        \\
							& ACA\hyp CPU       & N/A       & N/A        \\
						    & ACA+HOSVD\hyp GPU & 00:00:09  & 00:00:33   \\
\hline
\hline
\end{tabular}
}
\end{table}

The relative error of $\matvec{y}$ obtained with ACA+HOSVD\hyp GPU relative to Full\hyp CPU (ground truth) is shown on the right axis of Fig. 11 for the case of $5$ mm$^3$ voxel resolution and PWC basis functions. The plot shows how the error changes as a function of the tolerance ($1e-3$, $\dots$, $1e-8$) used for ACA. In particular, the relative error remained approximately an order of magnitude higher than ACA's tolerance. Fig. 11 also shows plots for the ACA rank and the maximum Tucker rank of $Z_{\rm bc}$ (values on the left axis). Both ranks increased as the tolerance of ACA was decreased. We expect similar results for the other cases, but we were unable to assemble the full coupling matrix, due to its vast memory footprint.

\renewcommand{\thefigure}{11}
\begin{figure}[ht!]
\begin{center}
\centering
\includegraphics[width=0.48\textwidth, trim={0 0 0 0}]{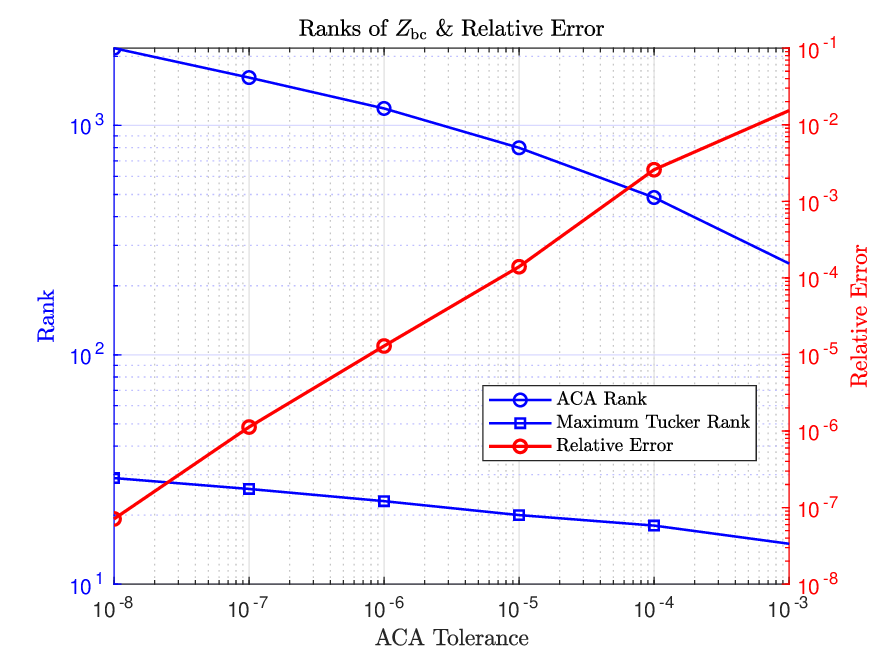}	
\caption{(left axis) ACA rank and maximum Tucker rank (obtained with HOSVD) of $Z_{\rm bc}$. (right axis) Error in $\matvec{y} = Z_{\rm bc} \matvec{x}$ calculated with ACA+HOSVD\hyp GPU relative to Full\hyp CPU.} 
\label{fig: n11}
\end{center}
\end{figure}

\section{Discussion} \label{sc:V}      

We showed that our new Tucker\hyp based algorithms could effectively compress the coupling VSIE matrices in the case of fine voxel resolutions. Thanks to the achieved compression factors, the matrix\hyp vector products can be performed in GPU's, yielding faster execution. The proposed approach will be key for a VSIE\hyp based in vivo implementation of GMT, for which reducing memory requirements and computation time are both critical factors. 
\par
For cases in which the coil is placed at some distance from the imaging object, the coupling matrix is low\hyp rank, thus can be compressed using ACA as $UV^*$. For coarse voxel resolutions, such compression strategy alone is effective and allows for a rapid CPU implementation of matrix\hyp vector products. However, as the voxel grid of the VIE domain is refined, the memory demand of $U$ increases until the standard ACA becomes impractical for applications that need high accuracy, such as GMT. In fact, in order to keep using ACA, one would have to relax the tolerance, sacrificing accuracy. To avoid this, in this work we introduced an extension of ACA (Algorithm 3), for which the memory demand remains as a low as the memory required by one column of the coupling matrix for any tolerance. Furthermore, Algorithm 3 can be executed in GPU for rapid computations also in the case of fine voxel resolutions. 
\par
An important aspect of the approach presented in this manuscript is that it can effectively compress the coupling matrix both when the coil is close to or far from the scatterer. Because of this, our method allows using GPUs to accelerate the matrix\hyp vector products for most applications. For example, if the close-fitting coil geometry in Fig. 7 were integrated with a head gradient insert with a surrounding conductive shield at a certain distance, with our approach the coupling matrix would still be compressible and fit in the memory of a GPU. In fact, the interactions between the shield and the head would have lower Tucker ranks than the ones between the coil's conductors and the head (see Section IV.A.1). On the other hand, the pFFT+Tucker method would no longer be efficient because it would require extending the VIE domain to fully include the shield. Even though the Green's function tensors of the extended domain would still be compressible with Tucker decomposition, the unknowns that multiply such tensors element-wise would not be. In fact, their dimensions would have to significantly increase in order to perform the relevant FFT\hyp based matrix\hyp vector products. 
\par
For the previous example, in order to fully exploit the highly parallel architecture of the GPU with a minimal number of matrix\hyp vector products operations, one could use a hybrid method that combines pFFT and Algorithm 3. To do that, the near and far interactions between the VIE domain and the conducting surfaces would need to be separated. Then, the pFFT could be used to model the near interactions between the VIE domain and the coil as in \cite{guryev2019fast}, while the arising Green's function operators could be compressed with the Tucker decomposition as in \cite{giannakopoulos2019memory}. Finally, the remaining far interactions between the VIE domain and the coil could be modeled with a coupling matrix, which would be vastly compressed with Algorithm 3. Such hybrid method could enable us to rapidly execute the matrix\hyp vector products in GPU for most cases, including complex coil-shield geometries, fine voxel resolutions, and PWL basis functions. The described hybrid method will be investigated in future work. 
\par
The main limitation of our proposed method is that it requires a considerable amount of time for the assembly of the coupling matrix, when the matrix is not compressible with ACA. It is especially slow because each element of the coupling matrix is a 5D integral. We showed that this could be addressed by implementing matrix assembly and compression in parallel, but such approach is not always possible due to memory limitations (IV.B.1). For such cases, one could alternatively employ 3D cross\hyp Tucker approximation algorithms \cite{Oseledets2008, Giannakopoulos2018}, which are less efficient than HOSVD for the tensor dimensions in this work, but do not suffer from memory constraints in case of large tensors. In fact, 3D cross\hyp Tucker methods require only a small number of rows, columns, and fibers of the tensor they approximate, and they can be implemented with linear complexity (with respect to tensor's linear size). In future work, we will explore the execution of multiple 3D cross-Tucker steps in parallel to avoid memory overflows when assembling a compressed coupling matrix in the case of extremely fine resolutions. Furthermore, the application of the tensor train decomposition \cite{Oseledets2011} on 4D reshapes of the coupling matrix will be investigated for remote geometries like the one appearing in section IV.B.2, as an alternative to Algorithm 3. 

\section{Conclusion} \label{sc:VI}

We presented a memory compression technique for the coupling matrix in VSIE systems. Our method enables one to form and store the coupling matrix even when its full form size is prohibitively large ($\sim 80$ TB). Specifically, in this work we were able to achieve a compression between $\sim 0.5$ (PWC) and $\sim 2$ (PWL) million times when simulating interactions between MRI coils and realistic body models with some distance between them, in the case of fine voxel resolutions of $1$ mm$^3$. The error was around one order of magnitude higher than the tolerance employed for the algorithm. The stored, compressed matrices could be used multiple times without the need to repeat the assembly. For example, this would allow one to rapidly perform EM simulations for the same coil with different body geometries, as far as they are contained in the original computational domain. For most cases, our compression method enables fitting large coupling matrices in GPUs, resulting in rapid execution of the VSIE matrix\hyp vector product (from $1$ to $56$ seconds for the studied scenarios). Finally, the proposed method could facilitate the implementation of VSIE\hyp based GMT for in vivo mapping of tissue electrical properties at clinically\hyp relevant voxel resolutions. 

\Urlmuskip=0mu plus 1mu\relax
\bibliographystyle{IEEEtran}
\bibliography{IEEEabrv,References}

\end{document}